\documentclass[12pt,a4paper]{iopart}
\pdfoutput=1 \widowpenalty10000 \clubpenalty10000
\usepackage[T1]{fontenc}\usepackage[latin1]{inputenc}
\usepackage{graphicx,color,booktabs,wrapfig,microtype,sidecap,floatflt,afterpage,cite,bm,hvfloat} \graphicspath{{Figures/}}
\usepackage[slantedGreek]{mathptmx}

\makeatletter
\long\def\@makecaption#1#2{\vskip\abovecaptionskip{\bf #1.} #2\vskip\belowcaptionskip}
\makeatother

\usepackage{hyperref}
\hypersetup{colorlinks, plainpages=false, linkcolor=blue, urlcolor=blue, citecolor=blue, pdfpagemode=UseNone, pdfstartview=FitBH}

\begin{document}

\makeatletter
\def\@evenhead{\hfil\itshape\rightmark}
\def\@oddhead{\itshape\leftmark\hfil}
\renewcommand{\@evenfoot}{\hfill\small\raisebox{-1em}{--~\textbf{\thepage}~--}\hfill}
\renewcommand{\@oddfoot}{\hfill\small\raisebox{-1em}{--~\textbf{\thepage}~--}\hfill}
\makeatother

\title[A.\,A.~Kulbakov \textit{et al.} \hfill Stripe-\textit{yz} magnetic order in the triangular-lattice antiferromagnet KCeS$_2$]{\vspace{-5em}\LARGE\\Stripe-\textit{yz} magnetic order in the triangular-lattice antiferromagnet KCeS$_2$}

\author{Anton~A. Kulbakov,$\!^{1,2}$ Stanislav~M. Avdoshenko,$\!^{3}$ In\'{e}s Puente-Orench,$\!^{4,5}$ Mahmoud Deeb,$\!^{1}$ Mathias Doerr,$\!^{1}$ Philipp Schlender,$\!^{6}$ Thomas Doert,$\!^{6}$ and Dmytro~S.~Inosov$^{1,2}$}

\address{\begin{tabular}{@{}r@{\,}p{0.845\textwidth}}
$^1$&Institut f\"ur Festk\"orper- und Materialphysik, Technische Universit\"at Dresden, 01069 Dresden, Germany.\\
$^2$&W\"urzburg-Dresden Cluster of Excellence on Complexity and Topology in Quantum Matter\,---\,\textit{ct.qmat}, TU~Dresden, 01069 Dresden, Germany.\\
$^3$&Leibniz-Institut f\"ur Festk\"orper- und Werkstoffforschung (IFW Dresden), Helmholtzstra{\ss}e 20, 01069 Dresden, Germany.\\
$^4$&Instituto de Nanociencia y Materiales de Arag\'on (INMA), CSIC-Universidad de Zaragoza, Zaragoza 50009, Spain.\\
$^5$&Institut Laue-Langevin, 71 avenue des Martyrs, CS 20156, 38042 Grenoble CEDEX 9, France.\\
$^6$&Fakult\"at f\"ur Chemie und Lebensmittelchemie, Technische Universit\"at Dresden, 01062 Dresden, Germany.\\
\end{tabular}}

\ead{\href{mailto:dmytro.inosov@tu-dresden.de}{dmytro.inosov@tu-dresden.de}}

\begin{abstract}
\noindent \hspace*{-1ex} Yb- and Ce-based delafossites were recently identified as effective spin-1/2 antiferromagnets on the triangular lattice. Several Yb-based systems, such as NaYbO$_2$, NaYbS$_2$, and NaYbSe$_2$, exhibit no long-range order down to the lowest measured temperatures and therefore serve as putative candidates for the realization of a quantum spin liquid. However, their isostructural Ce-based counterpart KCeS$_2$ exhibits magnetic order below $T_{\rm N}=400$~mK, which was so far identified only in thermodynamic measurements. Here we reveal the magnetic structure of this long-range ordered phase using magnetic neutron diffraction. We show that it represents the so-called ``stripe-$yz$'' type of antiferromagnetic order with spins lying approximately in the triangular-lattice planes orthogonal to the nearest-neighbor Ce--Ce bonds. No structural lattice distortions are revealed below $T_{\rm N}$, indicating that the triangular lattice of Ce$^{3+}$ ions remains geometrically perfect down to the lowest temperatures. We propose an effective Hamiltonian for KCeS$_2$, based on a fit to the results of \textit{ab initio} calculations, and demonstrate that its magnetic ground state matches the experimental spin structure.
\vspace{-1em}\end{abstract}


\noindent\rule{\textwidth}{.7pt}

\section{Introduction}
		
In low-dimensional quantum magnets, competing exchange interactions may lead to a strong frustration accompanied by enhanced quantum fluctuations. Ultimately this can prevent the systems from long-range order, and the ground state is supposed to be a quantum spin liquid (QSL)~\cite{Lacroix2011, Ramirez1994, Greedan2001, KnolleMoessner19}. The theoretically predicted QSL state remains experimentally elusive, and a major ongoing research effort is directed towards the identification of promising QSL candidate materials~\cite{BroholmCava20}.

Two most promising search directions are pursued. The first is motivated by the seminal work of Kitaev~\cite{Kitaev06} and is focused on materials with anisotropic Kitaev interactions on tricoordinated lattices~\cite{ChaloupkaJackeli10, SinghManni12, YamajiNomura14, AlpichshevMahmood15, KitagawaTakayama18, TakayamaKato15, ModicSmidt14, SandilandsTian15, NasuKnolle16, BanerjeeBridges16, BanerjeeYan17, Trebst17, GordonCatuneanu19, TakagiTakayama19}. There is no requirement of geometrical frustration in this case, because the frustration originates from highly anisotropic nearest-neighbor interactions that can result from strong spin-orbit coupling. The most promising candidate compounds in this class are honeycomb-lattice iridates ($\alpha$-Li$_2$IrO$_3$, Na$_2$IrO$_3$, H$_3$LiIr$_2$O$_6$)~\cite{ChaloupkaJackeli10, SinghManni12, YamajiNomura14, AlpichshevMahmood15, KitagawaTakayama18} or ruthenates ($\alpha$-RuCl$_3$)~\cite{SandilandsTian15, NasuKnolle16, BanerjeeBridges16, BanerjeeYan17, Trebst17} and their three-dimensional polymorphs that form hyper-honeycomb ($\beta$-Li$_2$IrO$_3$)~\cite{TakayamaKato15} or stripy-honeycomb ($\gamma$-Li$_2$IrO$_3$) \cite{ModicSmidt14} structures. The difficulty with these model systems is that in real materials, Kitaev interactions coexist with Heisenberg exchange and other possible types of interactions, leading to very intricate magnetic Hamiltonians~\cite{RauLee14, TakagiTakayama19, RusnackoGotfryd19}. Existing materials usually end up in different regions of the multidimensional parameter space away from the relatively small theoretically predicted stability regions of the QSL state.

The second search direction rests on the original proposal by Anderson~\cite{Anderson73} that a QSL ground state can be realized in two-dimensional (2D) spin-1/2 triangular-lattice antiferromagnets (TLAF) as a result of geometrical frustration~\cite{Balents10, Li2020}. The same is true also for other highly frustrated antiferromagnetic (AFM) 2D lattices, such as the kagome lattice~\cite{Sachdev92}, which is realized in many copper-containing compounds where quantum spins $S=1/2$ reside on the magnetic Cu$^{2+}$ ions, most famously in herbertsmithite~\cite{HermeleRan08, HanHelton12, HanNorman16, Norman16}. The difficulty with these systems, however, is that the lattice may experience small structural distortions which are often sufficient to relieve the frustration~\cite{Norman16, Inosov18}. Moreover, to the best of our knowledge, no realizations of an undistorted AFM triangular lattice of Cu$^{2+}$ spins have been identified to date among inorganic compounds. Therefore, the most suitable materials for the experimental verification of theoretical spin-1/2 models on the triangular lattice are layered rare-earth compounds with Ce$^{3+}$ or Yb$^{3+}$ ions, characterized by the 4\textit{f}$^1$ and 4\textit{f}$^{13}$ electronic configurations, respectively. Whenever the ground-state doublet is sufficiently separated from higher-energy states due to the crystal electric field (CEF) splitting, these ions realize an effective $\tilde{S}=1/2$ state at low temperatures. The most studied QSL candidate from this class is YbMgGaO$_4$~\cite{Li2015, Li2015a, Paddison2017, Kimchi2018}, although most recent studies tend to attribute the absence of long-range order in this material to Mg$^{2+\!}$-\,Ga$^{3+}$ site intermixing~\cite{LiAdroja17, ZhuMaksimov17}.

\begin{wrapfigure}[20]{r}{0.4\textwidth}\vspace{-1.6em}
\noindent\includegraphics[width=0.4\textwidth]{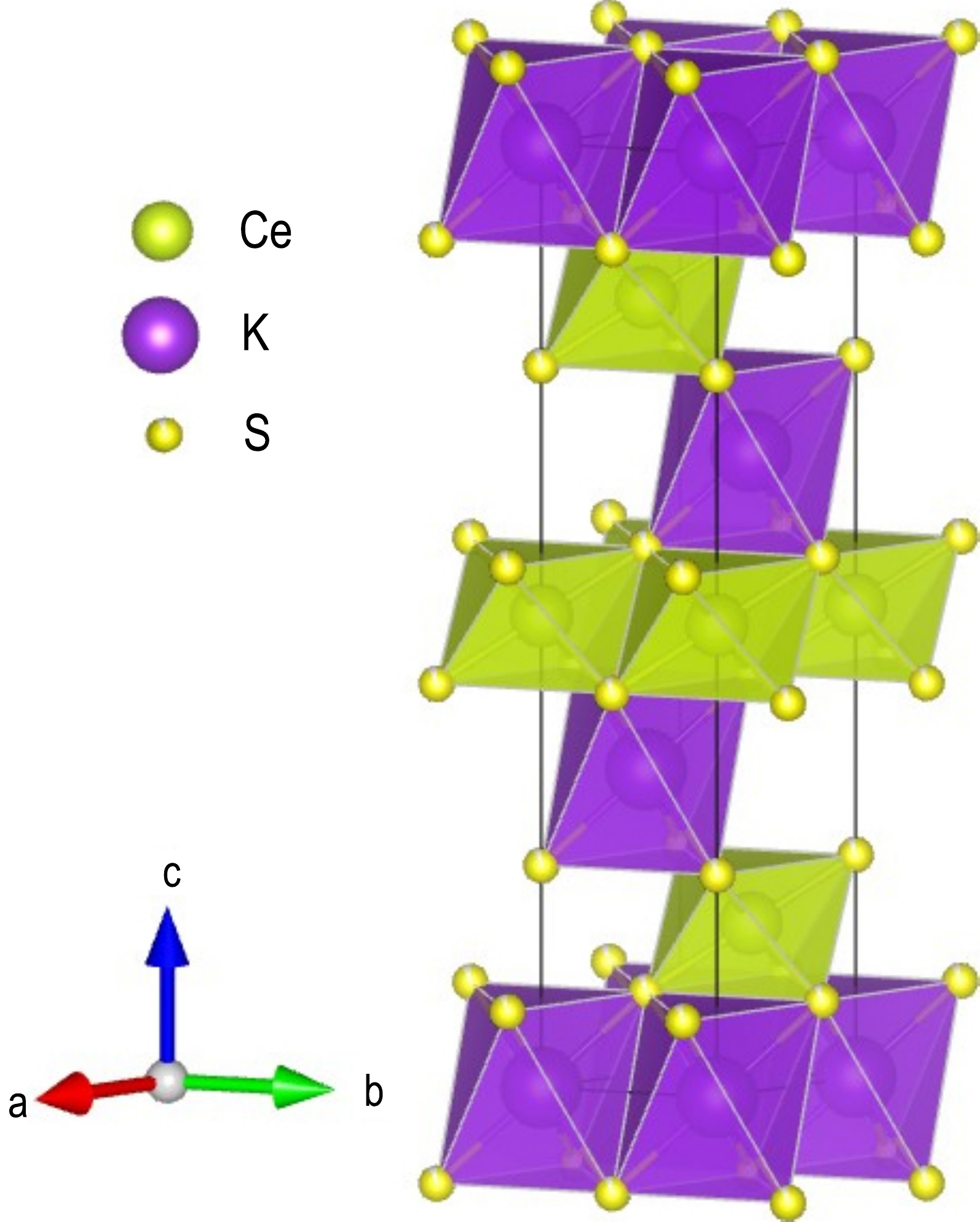}
\caption{Crystal structure of the KCeS$_{2}$ delafossite. The visualization was done in VESTA \cite{Momma2011}.}
\label{Fig:Crystal_structure}
\end{wrapfigure}

A short while ago, a whole new family of compounds with the undistorted triangular lattice of rare-earth ions with little or no structural disorder has been identified~\cite{LiuZhang18, BaenitzSchlender18, XingSanjeewa20, SchmidtSichelschmidt21}. These are ternary compounds with the general chemical formula $A^{+}R^{3+}X^{2-}_2$, where $A$ is an alkali metal (e.g. Na, K, Rb, Cs), $R$ is a rare-earth ion (in our case, Ce or Yb), and $X$ is a chalcogen atom (O, S, or Se). They crystallize in the layered delafossite structure depicted in Fig.~\ref{Fig:Crystal_structure} that derives from the rock-salt structure upon cation ordering, realizing ABAB-type stacking of magnetic triangular-lattice layers~\cite{OhtaniHonjo87, KippVanderah90}. Low-temperature thermodynamic measurements reveal no signatures of magnetic ordering down to subkelvin temperatures in NaYbO$_2$~\cite{LiuZhang18, BordelonKenney19, DingManuel19, RanjithDmytriieva19, BordelonLiu20}, NaYbS$_2$~\cite{LiuZhang18, BaenitzSchlender18, Sarkar2019}, and NaYbSe$_2$~\cite{LiuZhang18, RanjithLuther19, DaiZhang21, ZhangLi21}, indicating that these Yb-delafossite compounds may possess QSL ground states. Long-range magnetic order in these systems can be achieved in a moderate magnetic field~\cite{RanjithDmytriieva19, RanjithLuther19, XingSanjeewa19}. On the other hand, their isostructural Ce-based counterpart KCeS$_2$ develops magnetic order below $\sim$400~mK already in zero field, according to the specific-heat data~\cite{Bastien2020}.

Because Ce$^{3+}$ and Yb$^{3+}$ ions have one electron or hole in the $f$ shell, respectively, and the CEF schemes of all the mentioned compounds have ground-state doublets that are well separated from the excited CEF states~\cite{BaenitzSchlender18, DingManuel19, BordelonLiu20, Bastien2020, DaiZhang21, ZhangMa21}, one would generally expect the spin-1/2 approximation to be equally applicable to these systems. However, they may still differ in terms of the magnetocrystalline anisotropy and have different eigenfunctions of the CEF ground state with respect to their decomposition in the $|J,J_z\rangle$ basis, which would place them in different regions of the parameter space of a generic effective Hamiltonian proposed for triangular-lattice antiferromagnets~\cite{ZhuMaksimov18, MaksimovZhu19}. To understand these differences, we have investigated the magnetic structure of the low-temperature ordered phase in KCeS$_2$ and its low-temperature lattice structure using powder neutron diffraction.

\vspace{-2pt}\section{Sample preparation}\vspace{-3pt}
\label{Sec:SamplePreparation}

The powder sample of KCeS$_2$ was prepared through the following synthesis route. Potassium carbonate and cerium dioxide were mixed in a 20:1 molar ratio and thoroughly ground in a porcelain mortar. A glassy carbon crucible was filled with this mixture and placed in a tube furnace. Before heating up to the target temperature, the whole apparatus including a reservoir for carbon disulfide was flushed with argon for 30~min. The mixture was heated up to 1050~$^\circ$C within 3~hours under an unloaded stream of argon (5~L/h). While dwelling one hour, a stream of argon of~0.5 L/h was used to carry
CS$_2$ into the hot zone to enable the sulfidization. Finally, the apparatus was allowed to cool down to 600~$^\circ$C within 6~hours and without further control down to ambient temperature under a slight argon stream. The ingot that formed in the glassy carbon crucible was then dissolved in water to set the insoluble KCeS$_2$ free. Filtering via paper and funnel and washing with water and ethanol produced the sample, mainly composed of intergrown crystals of the target compound and some Ce$_2$O$_2$S as a minority phase. The formation of oxidic by-products during the formation of rare-earth metal chalcogenides (and other compounds of rare-earth metals) is well known and can be attributed to the high oxygen susceptibility of the rare-earth metal ions. It can only be prevented in a completely oxygen-free setup. We also observed that mechanical manipulations (such as pressing, grinding, etc.) can foster decomposition of KCeS$_2$ with the formation of the Ce$_2$O$_2$S phase, therefore we avoided grinding the sample and used the as-synthesized powder in neutron-diffraction measurements.

\vspace{-2pt}\section{Powder neutron diffraction}\vspace{-3pt}
\label{Sec:NeutronDiffraction}
	
\begin{figure*}[!t]	
\includegraphics[width=\textwidth]{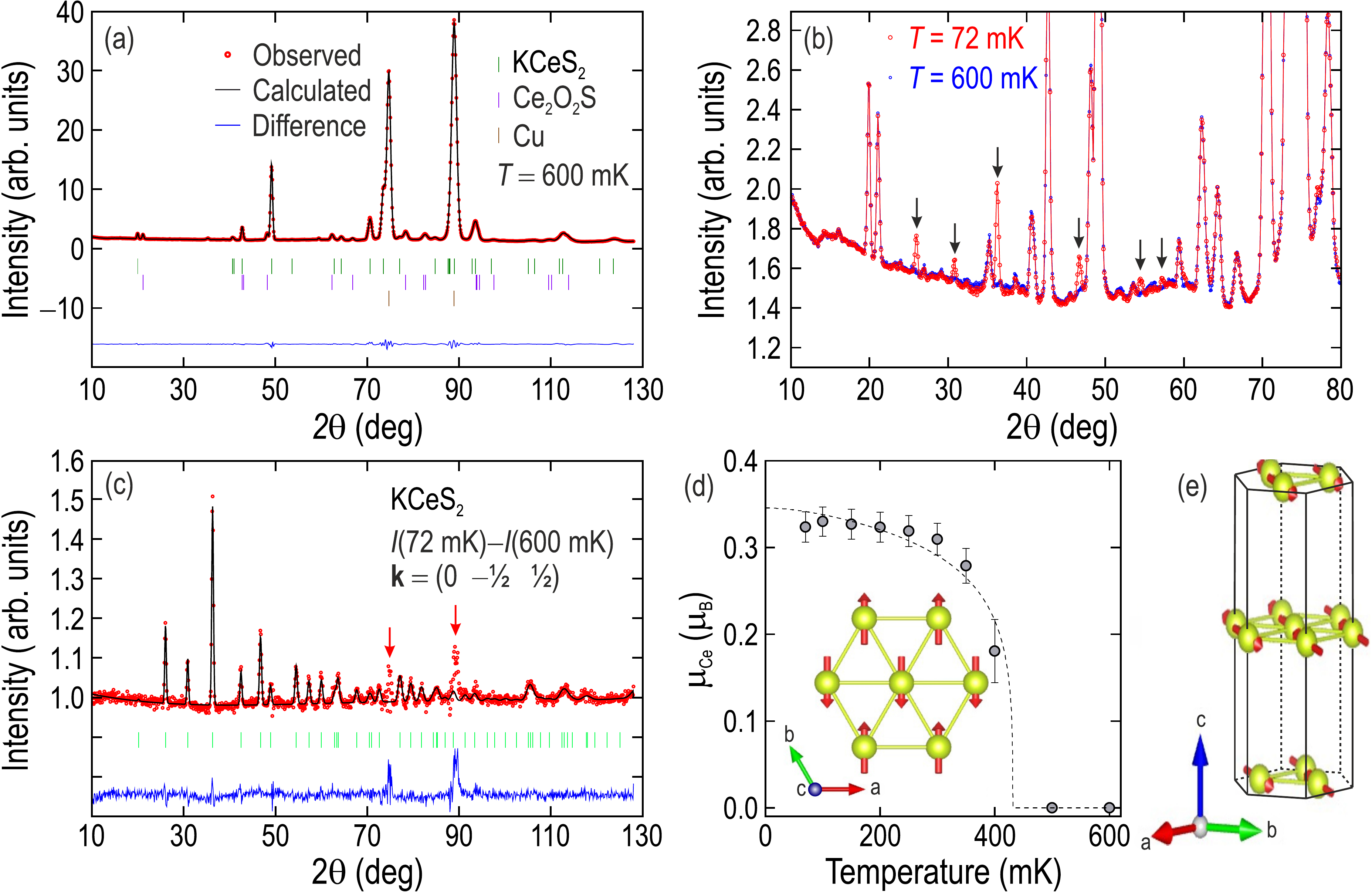}
\caption{Neutron powder diffraction data. (a)~Scattered neutron intensity at $T=600$~mK as a function of 2$\theta$, refined in the rhombohedral $R\overline{3}m$ space group. The fit includes Ce$_2$O$_2$S as an impurity phase and elemental Cu from the sample environment. Green, purple and brown marks denote peaks from KCeS$_2$, Ce$_2$O$_2$S, and Cu, respectively. (b)~Comparison of the scattered neutron intensity measured at $T=72$~mK~<~$T_{\rm N}$ (red) and $T=600$~mK~>~$T_{\rm N}$ (blue). Black arrows show the magnetic Bragg peaks, which appear below $T_{\rm N}$. (c)~The difference of intensities measured at low and high temperatures. Red arrows show the imperfect subtraction of strong Bragg reflections from Cu. (d)~Temperature dependence of the magnetic ordered moment. The refined magnetic structure in the $ab$ plane is shown in the inset. (e)~The resulting three-dimensional magnetic structure of KCeS$_2$.}
\label{Fig:Fit_data_D1B}		
\end{figure*}

To reveal the spin structure in the low-temperature AFM state of KCeS$_{2}$, we measured neutron powder diffraction at the D1B high-intensity two-axis powder diffractometer at ILL, France. Neutrons with a calibrated wavelength $\lambda = 2.5286$~\AA\ were selected with a highly oriented pyrolytic graphite (HOPG) (002) monochromator. The contribution of the instrument to the peaks broadening was determined from the refinement of Na$_2$Ca$_3$Al$_2$F$_{14}$ standard sample, while wavelengths were refined using a Si standard. Parasitic diffraction peaks arising from the sample environment were eliminated by a radial oscillating collimator. Our delafossite sample represented rough powder consisting of small crystallites with lateral sizes up to 0.5~mm. This sample was placed in a Cu can to ensure good thermal contact at subkelvin temperatures and was cooled down using a dilution refrigerator. The measurements were carried out while the sample was rotated about the vertical axis within 180$^{\circ}$ with a step of 1$^{\circ}$ (for the 72 and 600~mK datasets) or 3$^{\circ}$ (other temperatures), with subsequent averaging of the datasets to compensate for the preferred orientations of individual crystallites. The data were collected at several temperatures between 72 and 600~mK in zero magnetic field, allowing for sufficient time for temperature stabilization before every data collection. The measurements at base temperature and at 600~mK were measured with higher statistics, corresponding to approximately 13~h acquisition time per temperature for the 72 and 600~mK datasets.

The 600~mK dataset, collected above $T_{\rm N}$, is plotted in Fig.~\ref{Fig:Fit_data_D1B}\,(a). Apart from the structural reflections from the main KCeS$_{2}$ phase, it contains reflections from the paramagnetic impurity phase Ce$_2$O$_2$S~\cite{Zachariasen49, QuezelBallestracci70}, which constitutes about 15\% of our sample judging from the results of structural refinement but exhibits no magnetic Bragg reflections. There are also two intense peaks from metallic copper that originate from the sample container, which we included as the third phase. Our Rietveld refinement of the neutron data was done in the assumption of the rhombohedral $R\overline{3}m$ space group, previously reported for KCeS$_{2}$ from x-ray structure refinement~\cite{PlugVerschoor76, Bastien2020}. The agreement between the lattice structures obtained at room temperature and at 600~mK (apart from the changes in lattice constants due to thermal expansion) demonstrates the absence of any symmetry-lowering transitions down to $T_{\rm N}$. The low-temperature lattice parameters resulting from the refinement of our neutron data are summarized in Table~\ref{Tab:DiffParam}.

\begin{table*}[t!]\vspace{-1em}
\caption{\label{Tab:DiffParam} Low-temperature unit cell parameters, $\chi^2$ values, and $R$ factors for our refinement of neutron diffraction data on the majority phase KCeS$_2$ and minority phase Ce$_2$O$_2$S at $T=72$ and 600~mK. Here $V$ is the unit cell volume, $z({\rm S})$ is the $z$ position of S on the $6c$ Wyckoff site in KCeS$_2$, and $z({\rm Ce})$ and $z({\rm O})$ are the positions of Ce and O on the $4h$ Wyckoff site in Ce$_2$O$_2$S, respectively.\smallskip}
\centerline{
\begin{tabular}{@{~}l|l@{~}l|l@{~}l@{~}}
\toprule
\textbf{Parameters} & \multicolumn{2}{c|}{\bf KCeS$_{\rm\bf 2}$} & \multicolumn{2}{c}{\bf Ce$_{\rm\bf 2}$O$_{\rm\bf 2}$S} \\
\midrule
Fraction & \multicolumn{2}{c|}{84.3\%}  & \multicolumn{2}{c}{15.7\%} \\
Space group & \multicolumn{2}{c|}{$R\overline{3}m$ (\#166)} & \multicolumn{2}{c}{$P\overline{3}m1$ (\#162)} \\
\midrule
Temperature & \multicolumn{1}{c}{\bf 72~mK} & \multicolumn{1}{c|}{\bf 600~mK} & \multicolumn{1}{c}{\bf 72~mK} & \multicolumn{1}{c}{\bf 600~mK} \\
\midrule
$a$ (\AA) & ~~~~4.2242(1) & ~~~~4.2243(1) & ~~4.0011(3) & ~~4.0015(2) \\
$c$ (\AA) & ~~21.836(1) & ~~21.836(1) &  ~~6.888(1) & ~~6.888(1) \\
$V$ (\AA$^{3}$) & 337.44(2) & 337.46(2) &  95.50(1) & 95.52(1) \\
$z({\rm S})$ & ~~~~0.43(2) & ~~~~0.43(2) & \multicolumn{2}{c}{---} \\
$z({\rm Ce})$ & \multicolumn{2}{c|}{---} & ~~0.28(1) & ~~0.28(1) \\
$z({\rm O})$ & \multicolumn{2}{c|}{---}  & ~~0.62(1) & ~~0.62(1) \\
$\chi^{2}$ & ~~~~2.59 & ~~~~2.45 & ~~2.59 & ~~2.45 \\
$R_{\rm p} (\%)$ & ~~~~5.53 & ~~~~5.18 & ~~5.53  & ~~5.18  \\
$R_{\rm wp} (\%)$ & ~~~~5.66 & ~~~~5.53 & ~~5.66  & ~~5.53  \\
\bottomrule
\end{tabular}\vspace{-1em}
}
\end{table*}
	
The magnetic peaks, shown in the 72~mK dataset in Fig.~\ref{Fig:Fit_data_D1B}\,(b) with black arrows, set in below 400~mK\,---\,the transition temperature revealed earlier in specific-heat measurements~\cite{Bastien2020}. The difference of the low- and high-temperature datasets, plotted in Fig.~\ref{Fig:Fit_data_D1B}\,(c), contains only magnetic Bragg peaks, whereas the structural scattering from KCeS$_{2}$ and the oxysulphide impurity phase is subtracted to zero, which presents strong evidence for the absence of any structural phase transition associated with the magnetic ordering. The commensurate AFM propagation vector $(0\,\frac{\overline{1}}{2}\,\frac{1}{2})$ was determined by the K\_Search algorithm in the FullProf Suite based on difference of the two datasets. Representation analysis carried out with SARAh~\cite{Wills2000} for the $R\overline{3}m$ space group with the aforementioned propagation vector suggests two possible common irreducible representations: $\Gamma_2$ and $\Gamma_4$, which we have checked against the temperature-subtracted $I({\rm 0.072~K})-I({\rm 0.6~K})$ dataset. Our final result of the magnetic structure refinement, which was done using FullProf Suite~\cite{Rodrguez_Carvajal1993}, is shown in Fig.~\ref{Fig:Fit_data_D1B}\,(c) with a black line. The magnetic signal can be perfectly described only using the $\Gamma_2$ irreducible representation. Note that the two peaks marked with red arrows are of nonmagnetic origin, as they result from the imperfect subtraction of strong structural reflections from Cu. The magnetic structure of KCeS$_{2}$ is collinear, with spins lying approximately in the $ab$ plane and pointing orthogonally to the nearest-neighbor Ce-Ce bonds, as shown in the inset to Fig.~\ref{Fig:Fit_data_D1B}\,(d). The refinement suggests a statistically insignificant out-of-plane canting of Ce$^{3+}$ spins by $(5.5\pm5)^\circ$, yet fixing this angle to zero has no significant effect on the quality of the fit, so also the in-plane spin arrangement would be consistent with our data. This type of AFM structure on the triangular lattice is known as ``stripe-$yz$'' order among theorists~\cite{ZhuMaksimov18, MaksimovZhu19, SteinhardtMaksimov21, AversMaksimov21}. The stacking of the AFM layers along the out-of-plane direction is illustrated in Fig.~\ref{Fig:Fit_data_D1B}\,(e). Note that all magnetic Bragg peaks in Fig.~\ref{Fig:Fit_data_D1B}\,(c) are well described with our model for KCeS$_{2}$, suggesting that the impurity phase Ce$_2$O$_2$S develops no magnetic order down to 20~mK and could be therefore itself considered a promising spin-liquid candidate.

According to our combined refinement of the magnetic and structural data, the ordered magnetic moment on Ce$^{3+}$ at base temperature is approximately $0.32(1)\mu_{\rm B}$. It decreases monotonically with increasing temperature, as shown in Fig.~\ref{Fig:Fit_data_D1B}\,(d), following order-parameter behavior. Fitting this dependence with the empirical function $\mu(T)=A\,\tanh\left[\frac{\pi}{2}(T_{\rm N}/T-1)^\alpha\right]$, shown with a dotted line, gives a transition temperature $T_{\rm N} \approx 430$~mK, which is slightly higher than the value previously observed in thermodynamic measurements~\cite{Bastien2020}. The ordered moment in our data saturates below 300~mK, which may be a consequence of poor temperature stabilization in the powder sample close to the base temperature of the dilution refrigerator. Therefore, it cannot be excluded that we somewhat underestimate the value of the ordered moment at the base temperature.

\begin{wrapfigure}[26]{r}{0.532\textwidth}\vspace{-16pt}
\noindent\includegraphics[width=0.532\textwidth]{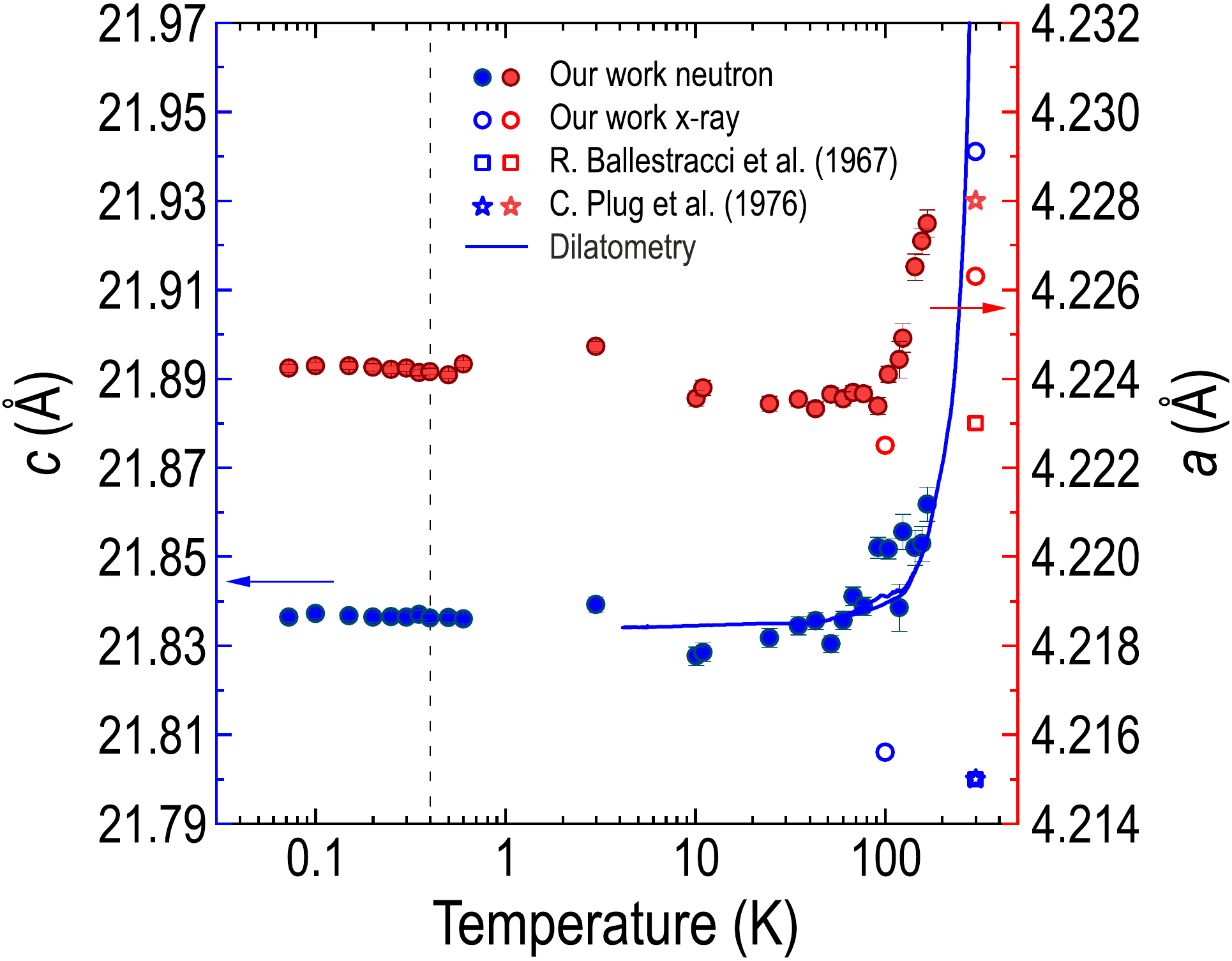}\vspace{-2pt}
\caption{Temperature dependence of the $a$ (red, right scale) and $c$ (blue, left scale) lattice parameters. Results of the structural refinement of neutron-diffraction data are shown with solid circles. The $c$-axis thermal-expansion measurement using capacitive dilatometry is overlayed with a solid line. We also show data points from our own and previously published~\cite{Ballestracci65, PlugVerschoor76} x-ray diffraction refinement of lattice constants with open symbols for comparison. The error bars are comparable with the point size. The vertical dashed line shows the N\'eel temperature. Note the logarithmic temperature scale.}
\label{Fig:T_a_c_dep}
\end{wrapfigure}

From the structural refinement of neutron diffraction data, we also obtained the temperature dependence of lattice constants, as it is shown in Fig.~\ref{Fig:T_a_c_dep} with solid circles. For comparison, we also show our own and previously published cell parameters~\cite{Ballestracci65, PlugVerschoor76} obtained using x-ray diffraction. Both lattice parameters $a$ and $c$ show conventional (positive) thermal expansion above 100~K, whereas below this temperature the thermal expansion is rather small and not measurable within the experimental error. We compare this result with the $c$-axis thermal expansion measured by capacitive dilatometry (solid line), carried out using our in-house dilatometer with a sensitivity to relative length changes of about $10^{-7}$~\cite{RotterMueller98}. The specimen consisted of several flakelike single crystals of KCeS$_{2}$ stacked along the $c$ axis to achieve a minimum required sample thickness. Therefore, only thermal expansion along the $c$ axis but not in the $ab$ plane could be measured. In good agreement with the neutron diffraction data, the capacitive dilatometry confirms a really small thermal expansion below 120~K, where the relative length change is only $3.5\times10^{-4}$. The linear expansion coefficient $\alpha$ stays less than $6\times10^{-6}$~K$^{-1}$ in this temperature range and only increases noticeably at higher temperatures. Additionally, nearly no magnetostriction could be detected during field sweeps (not shown), where the relative length change is only about $-1\times10^{-5}$ up to 8~T at $T=2$~K. The crossover in the behavior of the thermal expansion, observed consistently in diffraction and dilatometry, occurs more than two orders of magnitude higher in temperature than the magnetic ordering transition, which suggests that magnetostrictive effects become effective far above the N\'eel temperature due to the strong frustration in the system. Moreover, we observe no magnetostriction anomaly at the magnetic ordering temperature (dashed vertical line) in our neutron data. This is another indication that the AFM state is accompanied with no structural distortions, and the delafossite lattice structure with geometrically perfect triangular-lattice layers remains stable even after the stripe-$yz$ AFM order sets in.

\vspace{-2pt}\section{Theory}\vspace{-2pt}
\label{Sec:Theory}

\subsection{DFT/PBE/PAW level}

\begin{figure*}[b!]
\centerline{\includegraphics[width=0.8\textwidth]{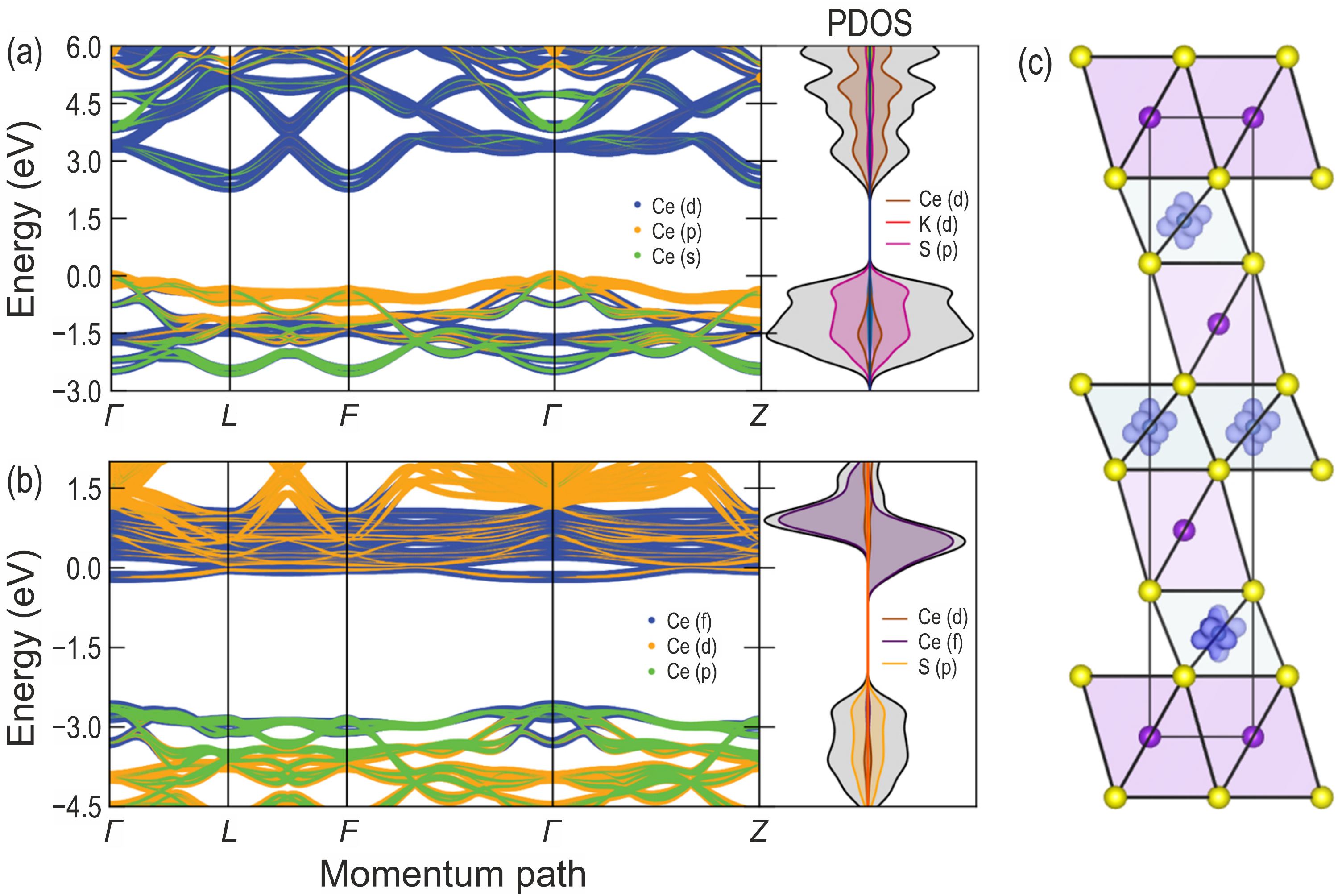}}
\caption{DFT/PBE/PAW bands structures and density of states of KCeS$_2$ with relevant orbital projections for the 4\textit{f} shell (a)~in the core and (b)~with explicit treatment of $4f$ electrons PAW potentials. (c)~The spin density ($\rho_\alpha - \rho_\beta$) for the spin-polarized solution.}
\label{Fig:Fig_Th1}		
\end{figure*}

We start the theoretical investigation at the density-functional theory (DFT/PBE/PAW) level using VASP suit\cite{KresseHafner93, PerdewErnzerhof96, KresseJoubert99}. Admittedly any single-determinant approximation would be a weak one for the 4\textit{f} system without special arrangements for $f$-shell. To deal with the complication due to the 4\textit{f} shell, at first the $f$ shell in core potential for Ce is considered \cite{KresseJoubert99}. The $\Gamma$-centered $\mathbf{k}$ grid of $4 \times 4 \times 2$ was used for system optimization and charge density convergence with plane-wave basis cutoff of 400~eV. The original rhombohedral structure has been optimized down to 10$^{-4}$ eV/\AA\ level with no restrictions on symmetry. The optimized structure preserves the original $R\overline{3}m$ space group but acquires a slightly expanded unit cell compared to the experimental data, $a_{\rm th}=4.23$~\AA\ vs. $a_{\rm exp}=4.22$~\AA\ and $c_{\rm th}=22.00$~\AA\ vs. $c_{\rm exp}=21.83$~\AA. While the lattice parameter $a$ remains within the experimental uncertainty from the measured value, the $c$ lattice constant experiences a significant (0.78\%) expansion along the $c$ axis, which accounts for $\sim$0.026~\AA\ expansion of the CeS$_6$ trigonal antiprism along the $C_3$ axis. Nevertheless, the electronic structure is only slightly affected by these differences. Figure~\ref{Fig:Fig_Th1} shows the band structure profiles along the high symmetry lines of KCeS$_2$ with 4\textit{f} shell in the core and explicit 4\textit{f}-shell treatment. Expectedly, with 4\textit{f} electrons in the core, KCeS$_2$ represents an indirect-gap semiconductor with large 2.7 and 2.3~eV direct and indirect gaps, respectively, as seen in Fig. \ref{Fig:Fig_Th1}\,(a). Explicit 4\textit{f}-shell treatment produces a localized state around the Fermi level with a small dispersion driven by \textit{d-f} hybridization. At the $\Gamma$ point, the \textit{d-f} hybridization is minimal with a formal gap of $\sim$0.05~eV (within the \textit{f} shell) and approximately 1.0~eV difference between the populated \textit{f} shell and the bottom of \textit{p-d} bands at the $\Gamma$ point. Without alteration of original per-site magnetization, the spin-polarized solution is converged to a ferromagnetic (FM) state as it is indicated by spin density (or magnetization density) in Fig.~\ref{Fig:Fig_Th1}\,(c).

\begin{figure*}[t]
\centerline{\includegraphics[width=0.8\textwidth]{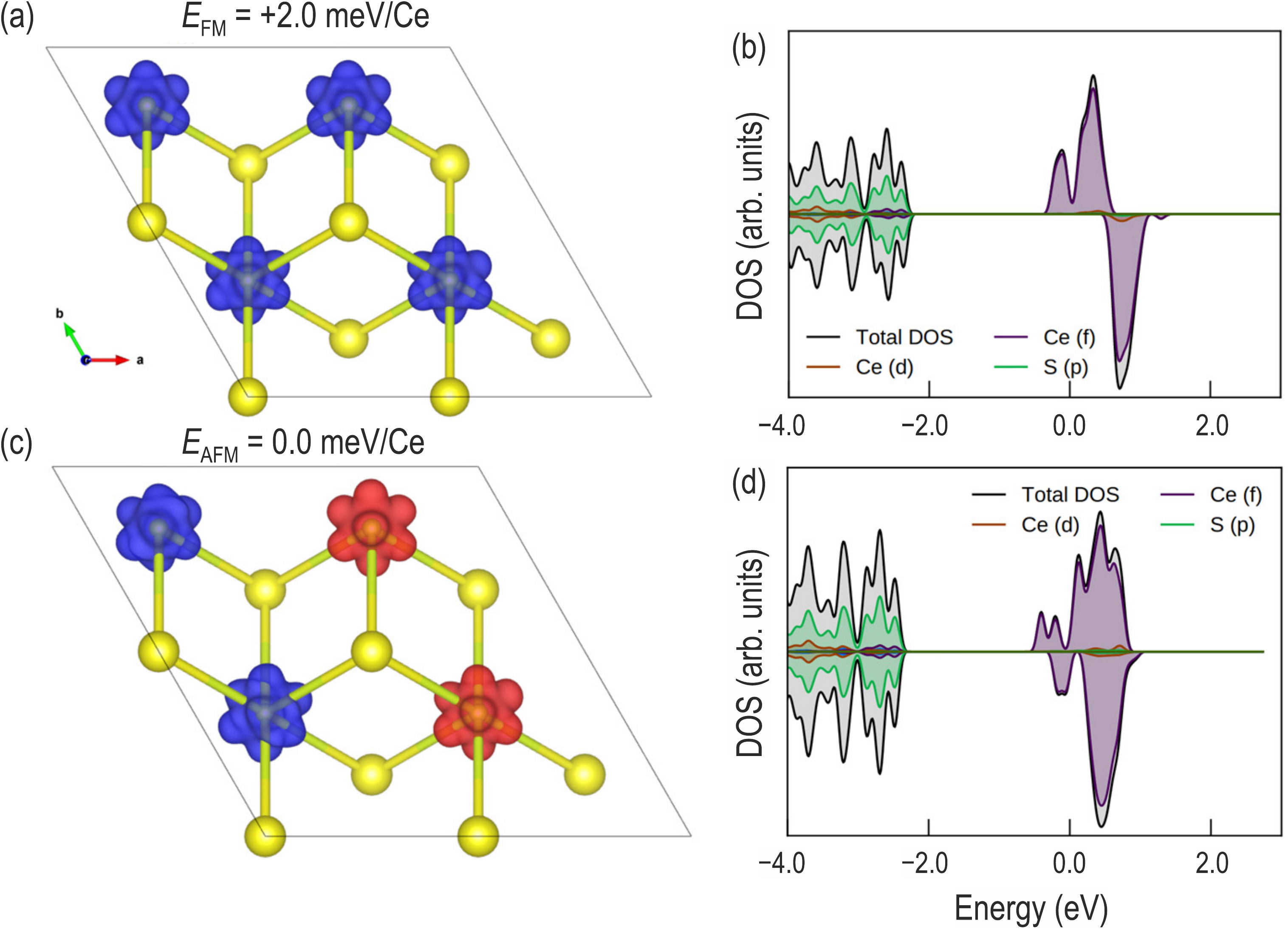}}
\caption{A 2$\times$2 supercell of a single CeS$_2$ layer with spin density isosurfaces and PDOS as predicted for spin-polarized DFT/PBE/PAW level with (a,b)~FM and (c,d)~AFM spin constraints.}
\label{Fig:Fig_Th2}		
\end{figure*}

Next, we estimate the magnetization energy, using the same DFT level as before, by optimizing a single-determinant wave function for different total magnetization densities: $\int(\rho_\alpha(r) - \rho_\beta(r)){\rm d}r=4$ for the FM state and $\int(\rho_\alpha(r) - \rho_\beta(r)){\rm d}r=0$ for the AFM state. Here, a proper consideration would require a supercell at least accommodating one propagation vector. However, the system of this size will suffer a lot more from well-known DFT shortcomings than a smaller system, and the question of applicability alone would need a separate study. Therefore, here we restrict the consideration to a fragment of a single isolated CeS$_2$ layer in a $2\times2$ supercell, as shown in Fig.~\ref{Fig:Fig_Th2}. We find that thermodynamically the AFM state is more stable than the FM state by 2.0~meV/Ce, as indicated in Figs.~\ref{Fig:Fig_Th2}\,(a,c). Similar to the complete system, the \textit{f} shells form condensed and localized bands around the Fermi level for both spin constraints [see Fig.~\ref{Fig:Fig_Th2}\,(b,d)]. Although the \textit{f} shell in the partial density of states (PDOS) has significant rearrangements, the spin density is well localized on the Ce sites and looks very similar before and after the spin flip. Further investigation of the moment orientation will require noncollinear consideration with a spin-orbit coupling Hamiltonian in place. This consideration would, even more, stretch the reliability of the DFT model in its application to \textit{f}-shell systems modeling. Moreover, 4\textit{f} systems have a strictly multiconfigurational character of the ground state $|J,m_J\rangle$-multiplet (with $\widehat{H}_{\rm SO} > \widehat{H}_{\rm LS}$). This cannot be easily accounted for with any DFT model. Multiconfigurational methods such as the complete active space self-consistent field (CASSCF) are generally inapplicable to a periodic system and limited to cluster approximations.

\subsection{CASSCF(1,7)/RASSI/SO modeling}

Luckily, we find our $f$ shell well isolated in the system. From band structures in Fig.~\ref{Fig:Fig_Th1}, we find that around the $\Gamma$ point, the 4\textit{f} states remain factorized from unoccupied states (lack of \textit{d}- or \textit{p}-projections and small $d$ contributions along the \textit{k} path). Also, the noticeable contribution of the $f$ shell to the valance bands is likely an artifact of \textit{d}+\textit{p} and $f$-shell projection procedure.

\begin{figure*}[b]
\centerline{\includegraphics[width=0.85\textwidth]{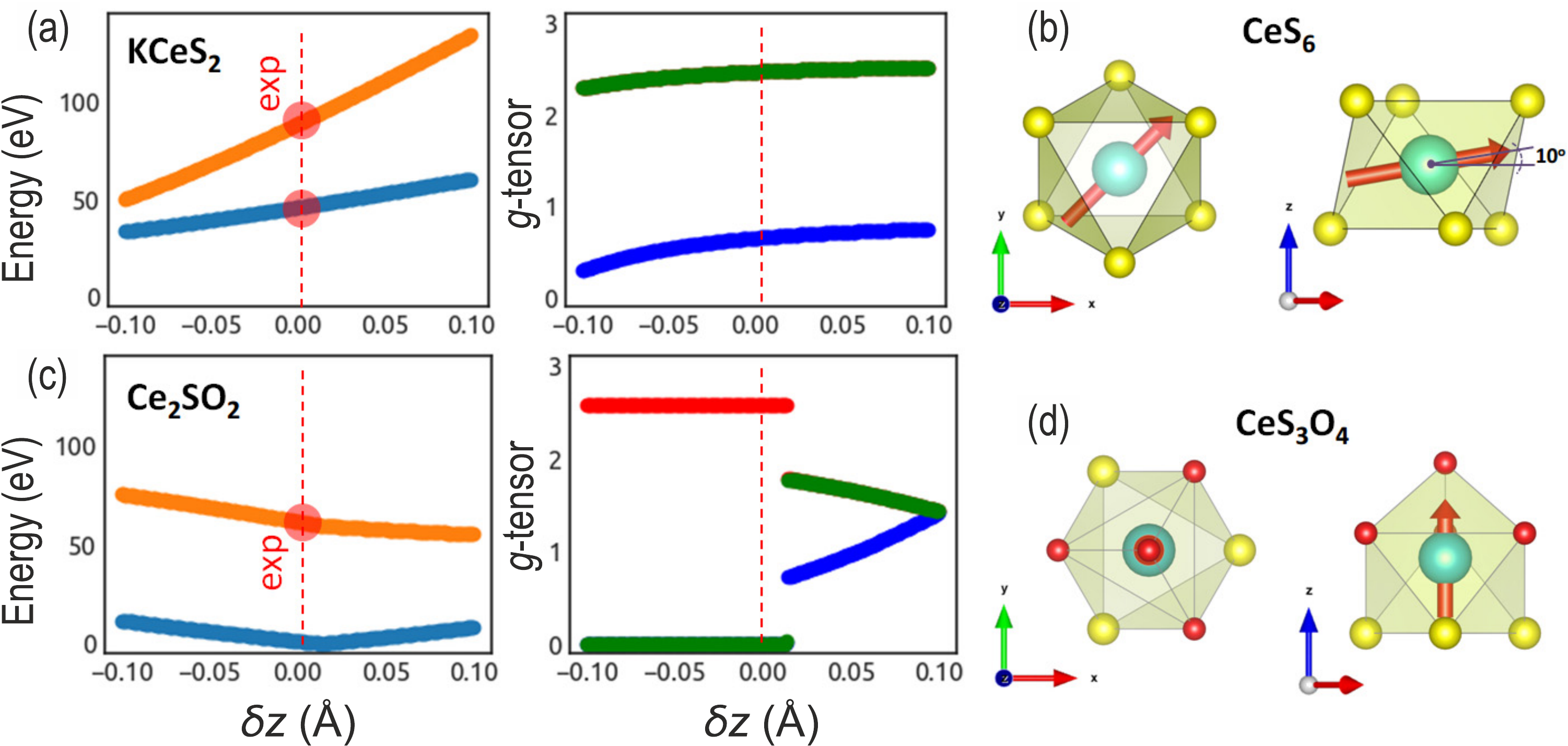}}
\caption{(a,c)~$^2F_{5/2}$-multiplet structure transformation (states energy and the ground-state $g$-tensor components) as a function of compression/elongation $\delta z$, defined relative to the experimental lattice structure of the (a)~CeS$_6$ and (c)~CeS$_3$O$_4$ polyhedra along the $C_3$ axis, obtained from point-charge modeling. (b,d)~The ground-state moment orientation in (b)~CeS$_6$ and (d)~CeS$_3$O$_4$ clusters.}
\label{Fig:Fig_Th3}		
\end{figure*}

Thus, to understand the local properties, we can restrict our consideration to the local polyhedron (minimal cluster of CeS$_6$) only. Moreover, as it was shown previously, the local properties of Ce centers in ESR spectra can be rationalized from very simple approximations \cite{Bastien2020}. Here, we elaborate more on this simple theoretical picture using the full strength of $^{\,2\!}F_{5/2}$ multiplet modeling based on Stevens parameters predicted through either point-charge models (using \textsc{McPhase}\cite{Rotter04} program and in-house Python scripts) or the CASSCF(1,7)/ANO-RCC-VDZP/RASSI/SO modeling (with \textsc{OpenMolcas} code\cite{AquilanteAutschbach20, ChibotaruUngur12}). We find that apart from some small scaling, both models provide virtually identical results and, for all intents and purposes, can be used interchangeably. We also apply the same level of theory to the impurity phase Ce$_2$O$_2$S for comparison and better understanding of the experimental results.

Figures \ref{Fig:Fig_Th3}\,(b,\,d) show two different projections of the CeS$_6$ and CeS$_3$O$_4$, respectively, based on the experimentally refined lattice parameters. In the case of CeS$_3$O$_4$ with an easy-axis anisotropy, the direction of the moment coincides with the direction of the main magnetic axis along $\mathbf{c}$. For the ``easy-plane'' situation in the CeS$_6$ cluster, the anisotropy ellipsoid and the radii are given by the $g$ tensor provided in Table~\ref{Table:Th1}. Still, it is possible to evaluate total angular momentum components along the main magnetic axis, $\boldsymbol{\mu}= \left(\mu_x=\left\langle 0\,|\,\mathbf{H}_{\rm Zee}\!\parallel\!\mathbf{x}\,|\,0\right\rangle=1.24;~\mu_y=\left\langle 0\,|\,\mathbf{H}_{\rm Zee}\!\parallel\!\mathbf{y}\,|\,0\right\rangle=1.24;~\mu_z=\left\langle 0\,|\,\mathbf{H}_{\rm Zee}\!\parallel\!\mathbf{z}\,|\,0\right\rangle =0.33\right)$. With no external perturbation, this moment has Larmor precession in a plane that is tilted $10^\circ$ with respect to the $ab$ plane, which is consistent with the moment canting suggested by the refinement of the magnetic structure from neutron diffraction in section~\ref{Sec:NeutronDiffraction}.

To gain more confidence in the employed theoretical framework, we first studied the impact of the compression/elongation along the $C_3$-axis of local Ce polyhedra (CeS$_6$ for KCeS$_2$ and CeS$_3$O$_4$ for Ce$_2$O$_2$S). Figure \ref{Fig:Fig_Th3} shows the multiplet transformation (state energies and $g$ tensor of the ground state) as a function of the elongation $\delta z$ defined relative to the experimentally refined crystal structure. Interestingly, drastic changes in the state $g$-energy diagram are observed within a small deviation of only $\pm 0.1$~\AA\ from the experimental value. For example, for KCeS$_2$ (CeS$_6$ cluster) the energies for the second and third states can change almost twofold and threefold, respectively, in this range. At the same time for Ce$_2$O$_2$S the situation is much more complex. Apart from the pronounced energy change, the system undergoes two local phase transitions from the easy-axis to easy-plane and back to easy-axis as $\delta z$ changes from 0.0 to 0.1. The detailed analysis of this phenomenon is outside the scope of this work, so we leave it for follow-up studies. In the following, we compare relative state energies around the experimental geometries and their compositions (Table~\ref{Table:Th1}).

\begin{table}
\caption{$^2F_{5/2}$ multiplet compositions, wave-function structures in the $|J, m_J\rangle$ basis, and $g$-tensor components for Ce$^{3+}$ in the CeS$_6$ and CeS$_3$O$_4$ polyhedra with experimental geometry. The last column gives calculated INS cross-sections for the corresponding CEF excitations.}
\centering
\begin{tabular}{c|c|c|lll|c}
\toprule
System             & Energy (meV) & Wave function & \multicolumn{1}{c}{$g_x$} & \multicolumn{1}{c}{$g_y$} & \multicolumn{1}{c|}{$g_z$} & INS, barn\\
\midrule
KCeS$_2$      & 0.00       & $0.96|\kern-1pt\pm\!1/2\rangle + 0.04|\kern-1pt\mp\!5/2\rangle$    & 2.47     & 2.47     & 0.65     & 1.91      \\
(CeS$_6$)     & 46.00      & $0.97|\kern-1pt\pm\!3/2\rangle + 0.03|\kern-1pt\mp\!3/2\rangle$     & 0.00     & 0.00     & 2.57     & 1.76      \\
                   & 88.98      & $0.96|\kern-1pt\pm\!5/2\rangle + 0.03|\kern-1pt\mp\!1/2\rangle$          & 0.1      & 0.1      & 4.08     & 0.23      \\
\midrule
Ce$_2$O$_2$S    & 0.00     &  $0.98|\kern-1pt\pm\!3/2\rangle + 0.02|\kern-1pt\mp\!3/2\rangle$         & 0.00    & 0.00     & 2.57     & 0.81      \\
(CeS$_3$O$_4$)  & 3.75       &    $0.70|\kern-1pt\pm\!1/2\rangle + 0.30|\kern-1pt\mp\!5/2\rangle$            & 1.818     & 1.818     & 0.650     & 1.29      \\
                         & 61.55      &   $0.70|\kern-1pt\pm\!5/2 \rangle + 0.30|\kern-1pt\mp\!1/2\rangle$ & 0.753     & 0.753     & 2.780     & 1.06      \\
\bottomrule
\end{tabular}
\label{Table:Th1}	
\end{table}

\begin{figure*}[t]\vspace{-2pt}
\mbox{
\begin{minipage}[b]{0.6\textwidth}
\includegraphics[width=\textwidth]{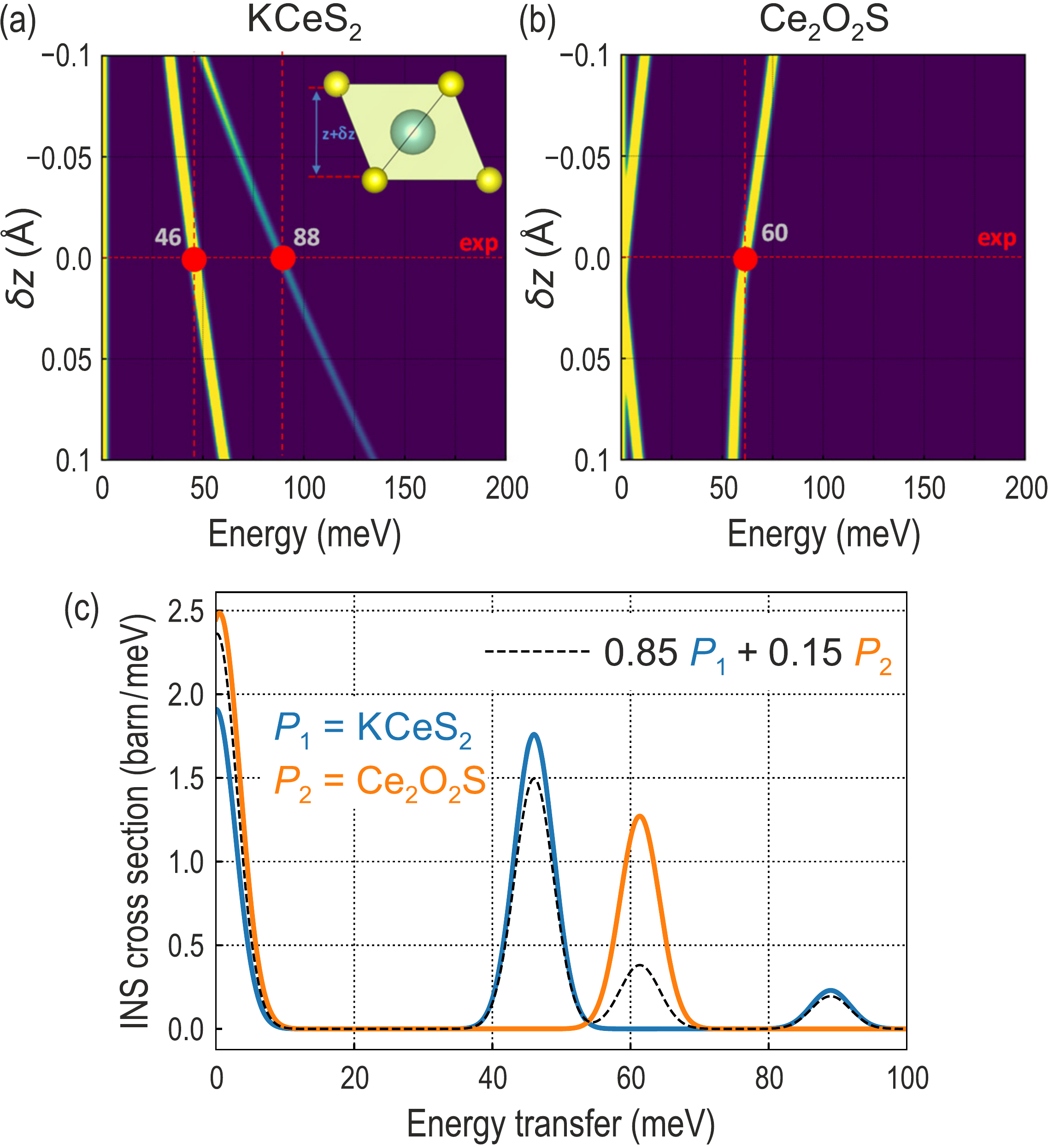}
\end{minipage}\hspace{-0.02\textwidth}
\begin{minipage}[b]{0.41\textwidth}
\caption{(a,b) 2D regular raster map of inelastic neutron scattering (INS) signals in energy-compression ($\delta z$) coordinates for KCeS$_2$ and Ce$_2$O$_2$S at $T=10$~K. (c)~INS spectra of individual KCeS$_2$ ($P_1$) and Ce$_2$O$_2$S ($P_2$) phases with experimental local polyhedron geometries ($\delta z = 0$) and an average spectrum weighted with experimental proportion between $P_1$ and $P_2$.}
\label{Fig:FigSI_th1}		
\end{minipage}}\vspace{-3pt}
\end{figure*}

In our previously published inelastic neutron scattering (INS) data~\cite{Bastien2020}, we observed one additional CEF excitation beyond those allowed for the Ce$^{3+}$ ion in KCeS$_2$. Such additional CEF lines may arise from an impurity phase \cite{BaenitzSchlender18}, from modified local environment due to lattice defects, such as stacking faults~\cite{GaudetSmith19, BordelonLiu20}, or from vibron quasibound states that result from strong magnetoelastic coupling~\cite{ThalmeierFulde82, AdrojaMoral12}. It is therefore important to verify whether the Ce$_2$O$_2$S impurity phase identified in our present work can account for this additional excitation. In the framework of the point-charge model, we have optimized the S-ion charge at $q_{\rm S}=-1.5e$, so that the first excited state of the $^2F_{5/2}$ multiplet matches the most intense experimental CEF transition at 46~meV, while using the experimental geometries for polyhedral CeS$_6$ and CeS$_3$O$_4$. In the case of CeS$_3$O$_4$, the O-ion point charge is assumed equal to that of the S ion, scaled by the ratio of Hirshfeld population ($P_{\rm O}/P_{\rm S}$) predicted by the DFT model: $q_{\rm O}=q_{\rm S}(P_{\rm O}/P_{\rm S})= -1.5e \cdot 5.0/3.7\approx-2.0e$. The calculated intensities of the INS cross section for CEF transitions in both compounds are calculated using the state energies and the wave functions and are given in Table~\ref{Table:Th1}. Taking the linear combination of the resulting spectra in the proportion KCeS$_2$\,:\,Ce$_2$O$_2$S = 85\,:\,15 (corresponding to the results of structural refinement given in Table~\ref{Tab:DiffParam}), we can reproduce both the energies and intensities of the experimental INS spectrum with remarkable accuracy (see Fig.~\ref{Fig:FigSI_th1}). This modeling suggests that the highest-energy peak at 74~meV originates from the KCeS$_2$ majority phase, despite its low intensity, whereas the additional peak at 62~meV can be assigned to the oxysulphide phase. As long as the charge of the ions changes consistently for both phases, the Ce$_2$O$_2$S excitation is always located between the two peaks originating from KCeS$_2$. This leaves no ambiguity in the assignment of the experimental CEF lines.

\vspace{-3pt}\subsection{The effective Hamiltonian parameterization}\vspace{-2pt}

To understand the origin of magnetic ordering in KCeS$_2$ and be able to compare it with rare-earth delafossites with a quantum-disordered ground state, we need to estimate effective interactions between local moments. The chosen DFT level predicts 2.0~meV/Ce, but this method is notoriously ill-equipped for such tasks. Thus, \textit{ab initio} verifications are needed.

For \textit{ab initio} estimations, however, one has to consider at least two interacting Ce$^{3+}$ ions at the same quantum-mechanical level. Fortunately, a pair of Ce$^{3+}$ atoms with 4\textit{f}$^{1}$ configuration each and an AFM ground state ($M=2S+1=1$) can be described by CASSCF(2,14)/ANO-RCC-VDZP/RASSI/SO problem with just 105 roots (Slater determinants) \cite{AquilanteAutschbach20}. This model produces the six low-lying spin orbitals within~60 meV. They comprise a single ground-state singlet, followed by a quasi-doublet at 5.1~meV, an excited singlet at~10.3 meV, and an excited quasi-doublet with an energy of 53.3~meV. For the single Ce$^{3+}$-ion, the \textit{ab-initio} ground-state multiplet structure can be reproduced by the minimal $|J,m_{J}\rangle$ basis using the derived \textit{ab-initio} Stevens parameters, $B^{\,k}_q$, with 99.98\% efficiency\cite{ChibotaruUngur12}. Assuming no significant electronic charge rearrangements (Ce$^{3+}$ sites are well-localized) and the Lines model as a valid approximation~\cite{Lines71}, we can project the two-site \textit{ab initio} solution on the following effective Hamiltonian in the $|J_1,m_{J_1}; J_2,m_{J_2}\rangle$ basis\vspace{-1pt}
\begin{equation}
\hat{\mathcal H}_{M} = \hat{\mathcal H}_{{\rm CF}_1} (B^{\,k}_q) + \hat{\mathcal H}_{{\rm CF}_2}(B^{\,k}_q)
-2\sum^{x,y,z}_{\alpha} J_{\alpha\alpha} \hat{J}_{\alpha,1} \hat{J}_{\alpha,2}
-2\sum^{x,y,z}_{\alpha\neq\beta} J_{\alpha\beta} \hat{J}_{\alpha,1} \hat{J}_{\beta,2}
\label{Eq:Th1}
\end{equation}\vspace{-4pt}
in the attempt to retrieve $J_{\alpha\beta}$ couplings.

\begin{table}
\caption{Effective interaction parameters of the Hamiltonian (\ref{Eq:Th1}), obtained by fitting to \textit{ab initio} CASSCF(2,14)/RASSI/SO states energies. The values are given in meV for two different moment models: $J=5/2$ and the dipolar moments $\langle I \rangle$ in \textsc{McPhase} \cite{Rotter04}.}
\centering
\begin{tabular}{cccccccccc}
\toprule
Moment model  & $J_{xx}$ & $J_{yy}$ & $J_{zz}$ & $J_{xy}$ & $J_{yx}$ & $J_{xz}$ & $J_{zx}$ & $J_{yz}$ & $J_{zy}$ \\ \midrule
5/2           & --0.70    & --0.60    & --4.90    & 0.0      & 0.0      & 0.0      & 0.0      & --0.8\,~     & --0.8\,~     \\ \midrule
$\langle I \rangle$ & --0.12    & --0.10    & --0.81    & 0.0      & 0.0      & 0.0      & 0.0      & --0.13    & --0.13    \\
\bottomrule
\end{tabular}\vspace{-2pt}
\label{Table:Th2}	
\end{table}

\begin{figure*}[b]\vspace{-2pt}
\mbox{
\begin{minipage}[b]{0.57\textwidth}
\includegraphics[width=\textwidth]{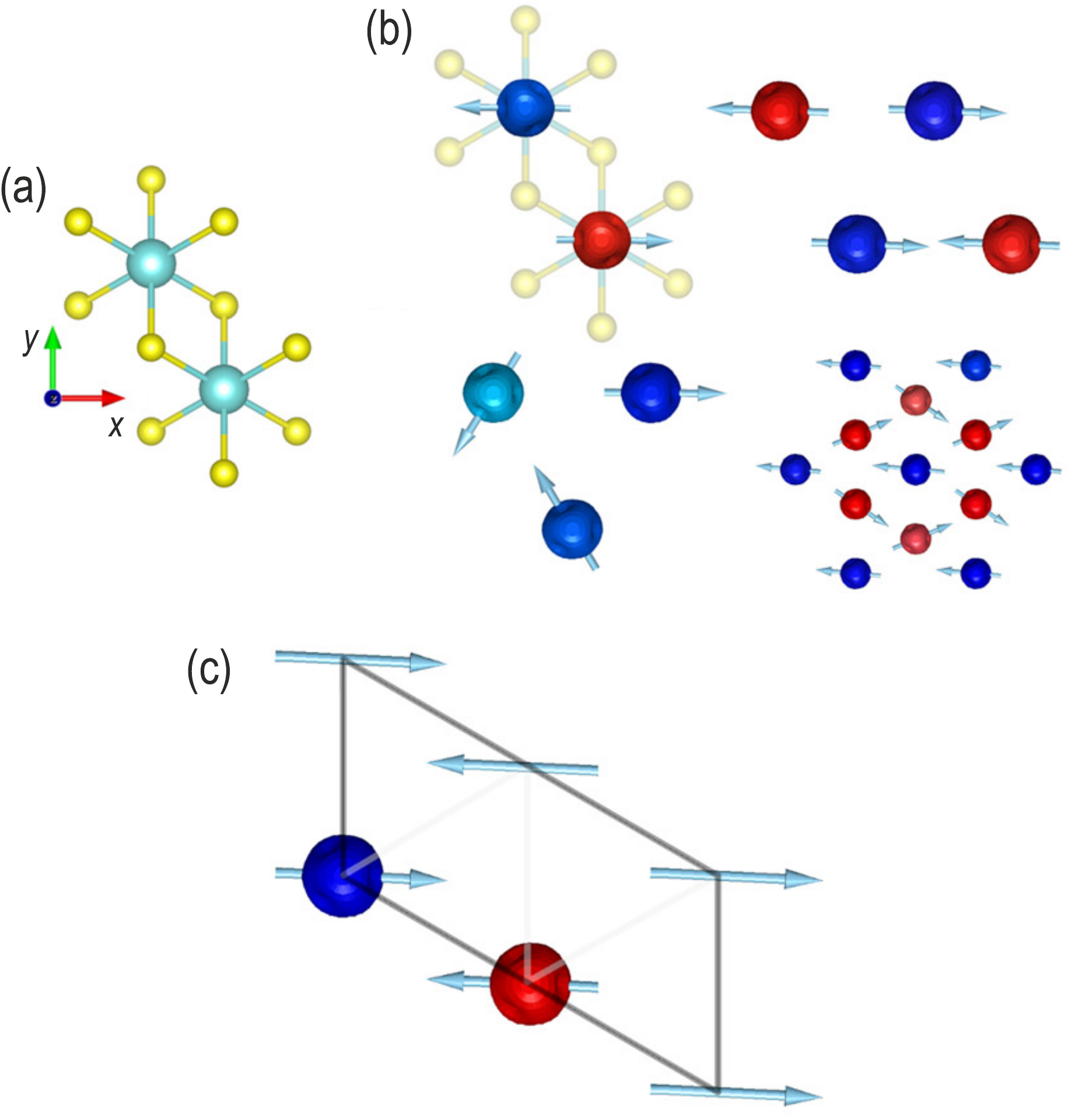}\vspace{1.2em}
\end{minipage}\hspace{-0.09\textwidth}
\begin{minipage}[b]{0.507\textwidth}
\caption{(a)~Geometry of the two-site model for \textit{ab initio} CASSCF(2,14)/RASSI/SO level calculations. (b) Magnetic order (orientation of the total moments) and on-site 4\textit{f} charge-density isosurfaces in clusters with 2, 3, 4, and 13 sites. (c)~Predicted magnetic unit cell in the CeS$_2$ layer and the couplings parameters estimated based in the \textit{ab initio} modeling of the two-sites system (see text for details).\vspace{-6pt}}
\label{Fig:FigSI_th2}		
\end{minipage}}
\end{figure*}

Using the Hamiltonian in Eq.~(\ref{Eq:Th1}), we fit the free parameters $J_{\alpha\beta}$ to \textit{ab initio} states energies predicted at CASSCF(2,14)/RASSI/SO. We find using the \textsc{PHI} program~\cite{ChiltonAnderson13} that the three lowest-energy states can be reproduced exactly with the optimized parameter values given in Table~\ref{Table:Th2}. Above 15.0~meV, the deviation in energy exceeds tens of meV, presumably due to contributions of the higher-order (multipolar) interactions for which the anisotropic Lines model, with dipolar interactions only, does not account. Moreover, the extracted Hamiltonian is based on the minimal two-site model which lacks the exact crystal symmetry and may overestimate local anisotropy contributions. Nonetheless, this model can describe the four low-lying states including the ground state. Thus, we conclude that this center symmetry model [Fig.~\ref{Fig:FigSI_th2}\,(a)] is able to predict noticeable anisotropy in interactions between local moments and sufficient to control total magnetic order in cluster models. To prove that, these interactions were further implemented in the localized-moments ordering models using the self-consistent mean-field method (see \textsc{McPhase}\cite{Rotter04} for more details) for clusters of 2, 3, 4, and 13 sites and for fully periodic systems with proper $q$-space sampling (Figure \ref{Fig:FigSI_th2}). Having allowed for the $\mathbf{q}$-sampling and a maximum unit cell up to 4 atoms in any direction, within the proposed interaction Hamiltonian, we obtain from this model the magnetic unit cell shown in Fig.~\ref{Fig:FigSI_th2}\,(f). This is consistent with the experimentally observed stripe-$yz$ order and the ordered moment of 0.33~$\mu_{\rm B}$/Ce, which also agrees with the experimental value.

\vspace{-2pt}\section{Discussion and conclusions}\vspace{-2pt}
\label{Sec:Discussion}

In this paper, we have presented the results of magnetic structure refinement for the low-temperature AFM state, which was recently revealed in the effective spin-1/2 triangular-lattice antiferromagnet KCeS$_2$. It represents collinear stripe-$yz$ AFM order with spins lying orthogonal to the nearest-neighbor Ce--Ce bonds in the $ab$ plane, possibly with a small ($\sim$10$^\circ$) out-of-plane canting of the magnetic moments that is also expected from theory. A similar stripe order was recently proposed for the closely related compound CeCd$_{3}$As$_{3}$ \cite{AversMaksimov21}. In addition, we have shown that there are no lattice distortions or symmetry-lowering structural transitions associated with the magnetic ordering to within $\sim$10$^{-4}$. The thermal expansion remains very small ($\alpha_c<6\!\times\!10^{-6}$~K$^{-1}$) below 120~K, which we confirmed for the $c$ lattice constant using capacitive dilatometry. Our experimental results also indicate that cerium oxysulphide, Ce$_2$O$_2$S, which was present in our sample as a minority phase, does not order magnetically down to 20~mK and may therefore represent a promising spin-liquid candidate deserving a separate study.

Following the experimental determination of the magnetic structure, we could explain it using first-principles calculations that predict exchange-coupling anisotropy in the pair interaction between two neighboring Ce$^{3+}$ ions. This resulting coupling tensor can be regarded as \textit{ab initio} because it was based on a direct fit of low-energy spectra of the exact RASSI/SO problem with no further approximations. We find that only a limited number of spin-orbitals can be used in the fit, which is direct evidence that at higher energy, coupled states are mediated by higher-order couplings beyond dipolar. We also find that, surprisingly, plain-vanilla DFT model at PBE/PAW level with explicit 4\textit{f}$^1$ treatment correctly predicts the dominance of AFM phase over FM with the reasonable magnetization density and the difference in energy between AFM and FM configurations of 2~meV/Ce, which is remarkably close to the mean isotropic coupling, $(J_{xx}+J_{yy}+J_{zz})/3=2.1$~meV/Ce.

Furthermore, comparing the ratio of $J_{ij}$ components in our model with the recent group-theoretical study with a pseudospin-1/2 moment model on a similar lattice \cite{MaksimovZhu19, SteinhardtMaksimov21}, we find that with $J_{yz}/J_{xx}\sim 1$ the KCeS$_2$ system should be deep in the stripe-$yz$ region of the phase diagram, in consistency with our experimental spin structure.

\vspace{-2pt}\section*{Acknowledgments}\vspace{-2pt}

This project was funded in part by the German Research Foundation (DFG) under the individual research grant IN~209/9-1, via the project C03 of the Collaborative Research Center SFB 1143 (project-id 247310070) at the TU Dresden, and the W\"urzburg-Dresden Cluster of Excellence on Complexity and Topology in Quantum Matter\,---\,\textit{ct.qmat} (EXC 2147, project-id 390858490). S.\,A. thanks A.~Popov (IFW, Dresden) and M.~Vojta (TU Dresden) for fruitful discussions and acknowledges financial support from the German Research Foundation (DFG) under Grant No. AV~169/3-1. We also acknowledge V.~Joyet and S.~Djellit for technical assistance and Institut Laue-Langevin, Grenoble (France) for providing neutron beam time~\cite{OnykiienkoInosovDOI}.

\vspace{-2pt}\section*{Bibliography}\vspace{-2pt}

\bibliographystyle{iopart-num}\enlargethispage{12pt}
\bibliography{KCeS2_diffraction}

\end{document}